\documentclass[conference]{IEEEtran}
\IEEEoverridecommandlockouts
\usepackage{cite}
\usepackage[caption=false,font=normalsize,labelfont=sf,textfont=sf]{subfig}
\usepackage{amsmath,amssymb,amsfonts}
\usepackage{algorithmic}
\usepackage{algorithm}
\usepackage{graphicx}
\usepackage{textcomp}
\usepackage{xcolor}
\usepackage{upgreek}
\usepackage{balance}
\usepackage{fancyhdr} 
\usepackage{multirow}
\usepackage{threeparttable}
\usepackage{cite} 

\def\BibTeX{{\rm B\kern-.05em{\sc i\kern-.025em b}\kern-.08em
    T\kern-.1667em\lower.7ex\hbox{E}\kern-.125emX}}

\begin{document}
\bibliographystyle{IEEEtran}

\title{Performance Evaluation of PAC Decoding with Deep Neural Networks\\
}
\author{
\IEEEauthorblockN{
Jingxin Dai\IEEEauthorrefmark{2}\IEEEauthorrefmark{1},
Hang Yin\IEEEauthorrefmark{2}\IEEEauthorrefmark{1},
Yansong Lv\IEEEauthorrefmark{2}\IEEEauthorrefmark{1},
Yuhuan Wang\IEEEauthorrefmark{2}\IEEEauthorrefmark{1}, and
Rui Lv\IEEEauthorrefmark{2}\IEEEauthorrefmark{1}}
\IEEEauthorblockA{\IEEEauthorrefmark{2}Engineering Research Center of Digital Audio and Video, Communication University of China, Beijing, China}
\IEEEauthorblockA{\IEEEauthorrefmark{1}State Key Laboratory of Media Convergence and Communication, Communication University of China, Beijing, China}
\IEEEauthorblockA{\{daijingxin, yinhang, lys\_communication, wangyuhuan, and lvrui\}@cuc.edu.cn}
\thanks{Hang Yin is the corresponding author (yinhang@cuc.edu.cn).}
}

\maketitle
\begin{abstract}
By concatenating a polar transform with a convolutional transform, polarization-adjusted convolutional (PAC) codes can reach the dispersion approximation bound in certain rate cases. However, the sequential decoding nature of traditional PAC decoding algorithms results in high decoding latency. 
Due to the parallel computing capability, deep neural network (DNN) decoders have emerged as a promising solution.
In this paper, we propose three types of DNN decoders for PAC codes: multi-layer perceptron (MLP), convolutional neural network (CNN), and recurrent neural network (RNN).
The performance of these DNN decoders is evaluated through extensive simulation. 
Numerical results show that the MLP decoder has the best error-correction performance under a similar model parameter number.
\end{abstract}

\begin{IEEEkeywords}
PAC codes, deep learning, neural network, decoder.
\end{IEEEkeywords}

\section{Introduction}   

Polar codes are the first channel codes that can achieve the capacity of any symmetrical binary input discrete memoryless channels (BI-DMCs) under the successive cancellation (SC) decoding algorithm \cite{1}. 
Nevertheless, the error-correction performance of finite-length polar codes under the SC decoding algorithm is not competitive. 
To address this issue, the SC list (SCL) decoding algorithm was introduced \cite{2}. 
When concatenated with cyclic redundancy check (CRC) codes \cite{3}, the CRC-aid SCL (CA-SCL) decoder yields significant improvements compared to standard polar codes. 
This advancement finds application in the enhanced mobile broadband (eMBB) scenario of 5G new radio (NR) systems \cite{4,5}. \par

Recently, a novel coding scheme named polarization-adjusted convolutional (PAC) codes was introduced, which concatenates a rate-1 convolutional pre-transformation with the polarization transform \cite{6}. 
Remarkably, the error-correction performance of PAC codes (128, 64) can reach the dispersion approximation (DA) \cite{7} under Fano sequential decoding. 
Furthermore, researchers presented the list decoding with the fixed complexity \cite{8,9}, which offers certain advantages over the Fano sequential decoding in terms of worst-case complexity \cite{9}. 
However, despite many fast list decoding algorithms \cite{10,11,12} being introduced to reduce the complexity, their sequential decoding nature still results in high latency and limited throughput.
This limitation hinders their practical application in communication systems, which require high reliability and low latency.\par

Deep learning (DL) \cite{13}, also known as deep neural network (DNN), provides a new direction to tackle this problem. 
Since its remarkable ability to tackle complex tasks, DL has facilitated significant advancements in various domains, including the communication physical layer \cite{14}.
Among all physical layer algorithms, since the general channel decoder can be treated as a classification task \cite{15}, it can be effectively implemented by DL.
The channel decoder based on DNNs is called the DNN decoder, which consists of two stages: training and testing. 
Two primary advantages of the DNN decoder are non-iterative and low-latency decoding. 
This is because the DNN decoder computes the estimated values of information bits by passing through each layer only once with the pre-trained neural network, which is referred to as one-shot decoding \cite{16}.
Compared to traditional decoding algorithms, DNN decoders can be quickly and efficiently implemented using the current DL platforms like PyTorch \cite{17}, along with powerful hardware such as graphical processing units (GPUs).\par

Researchers have explored the application of DNNs to solve the polar decoding problem. 
\cite{18} first presented the DNN decoder for polar codes, which has a similar correction-error performance of maximum a posteriori (MAP) decoding. 
Notably, \cite{18} pointed out that neural networks can learn a form of decoding algorithm, rather than only a simple classifier. 
The authors in \cite{19} introduced a low complexity belief propagation (BP) decoder with DNN, which can achieve around 0.25 dB gain at different code lengths. 
To meet the demand of various channel conditions, \cite{20} provided a DNN-based adaptive polar coding scheme. 
In response to the high average complexity of dynamic SCL flip (D-SCLF) decoding, deep learning-assisted adaptive D-SCLF decoding is introduced in \cite{21}.
Furthermore, \cite{22} presented a DNN-aided path-splitting selection strategy to enhance the efficiency of path splitting during SCL decoding, effectively avoiding unnecessary sorting operations without error-correction performance loss. \par

In this paper, we conduct a comprehensive performance evaluation of PAC decoders with different DNNs and analyze the features of each DNN decoder.
First, we propose three DNN decoders with similar parameter magnitude for PAC codes, which build upon multi-layer perceptron (MLP), convolution neural network (CNN), and recurrent neural network (RNN), respectively.
Then, we carry out extensive experiments to evaluate and analyze the performance of proposed DNN decoders. 
Simulation results show that the error-correction performance of the MLP decoder has a considerable advantage over its counterpart DNN decoder with a similar parameter number.

The remainder of this paper is organized as follows. 
Section II introduces the preliminaries of PAC codes and neural networks. 
Section III illustrates the system framework and proposed DNN decoders in detail. 
Simulation and numerical results of MLP, CNN, and RNN decoder are given in Section IV. 
Finally, Section V concludes the entire paper.

\section{Preliminaries} 

\textit{Notations}: Lowercase letters $(a, b, ...)$ are scalars, boldface lowercase letters $(\boldsymbol{a}, \boldsymbol{b}, ...)$ denote vectors, and boldface uppercase letters $(\boldsymbol{\rm{A}},\boldsymbol{\rm{B}}, ...)$ represent matrices. 
$\boldsymbol{b}[i]$ stands for the $i$-th element of vector $\boldsymbol{b}$. 
In addition, $\otimes$ indicates the Kronecker product.\par

\subsection{PAC Codes}

A PAC code can be represented by  $(N, K, {\mathcal{A}}, \boldsymbol{g})$, where $N$, $K$, $\mathcal{A}$, and $\boldsymbol{g}$ represent the codeword length, the information length, the set of information bit index, and the impulse response of convolution, respectively. 
Noted, the code lengths of PAC codes must be powers of two, i.e. $N=2^n$. Moreover, the code rate is $R=K/N$. \par

The PAC encoder consists of three steps. 
The first step is the rate-profiling, which maps the information vector $\boldsymbol{d}[1:K]$ to the bit sequence $\boldsymbol{v}[1:N]$ using  $\mathcal{A}$. 
The second step computes $\boldsymbol{u}[1:N]$ by the convolutional operation\par
\begin{equation}
\begin{aligned}
\boldsymbol{u}=\boldsymbol{v\rm{G}},
\end{aligned}
\end{equation}
where $\boldsymbol{\rm{G}}$ is an upper-triangular Toeplitz matrix denoting a rate-1 convolution operation and can be represented as
\begin{equation}
\begin{aligned}
\boldsymbol{\rm{G}} = \left[ {\begin{array}{*{20}{c}}
{\boldsymbol{g}[1]}&{\boldsymbol{g}[2]}&{...}&{\boldsymbol{g}[m]}&{...}&0\\
0&{\boldsymbol{g}[1]}&{\boldsymbol{g}[2]}& \ldots &{\boldsymbol{g}[m]}& \vdots \\
0&0& \ddots & \ddots & \ddots & \vdots \\
 \vdots & \ddots & \ddots & \ddots & \ddots & \vdots \\
 \vdots & \ddots & \ddots &0&{\boldsymbol{g}[1]}&{\boldsymbol{g}[2]}\\
 0 & \cdots & \cdots & \cdots &0&{\boldsymbol{g}[1]}
\end{array}} \right],
\end{aligned}
\end{equation}
where $\boldsymbol{g}[1]=\boldsymbol{g}[m]=1$, and $m$ is the constraint length of the convolution. During the third step, the PAC codeword $\boldsymbol{x}[1:N]$ is computed by the polar transformation
\begin{equation}
\begin{aligned}
\boldsymbol{x}=\boldsymbol{u\rm{F}}_N,
\end{aligned}
\end{equation}
where $\boldsymbol{\rm{F}}_N$ is the $n$-th Kronecker power of $\boldsymbol{\rm F}=[\begin{smallmatrix}1 & 0 \\ 1 & 1 \end{smallmatrix}]$, i.e., $\boldsymbol{\rm F}_N=\boldsymbol{\rm F}^{\otimes n}$. \par

\subsection{Neural Network}
In this paper, three effective neural network architectures (MLP, CNN, and RNN) are utilized to construct the DNN decoder. 
Therefore, we provide an overview of the fundamental architecture of each neural network. \par

\begin{figure}[!t]
\centering
\includegraphics[scale = 0.042]{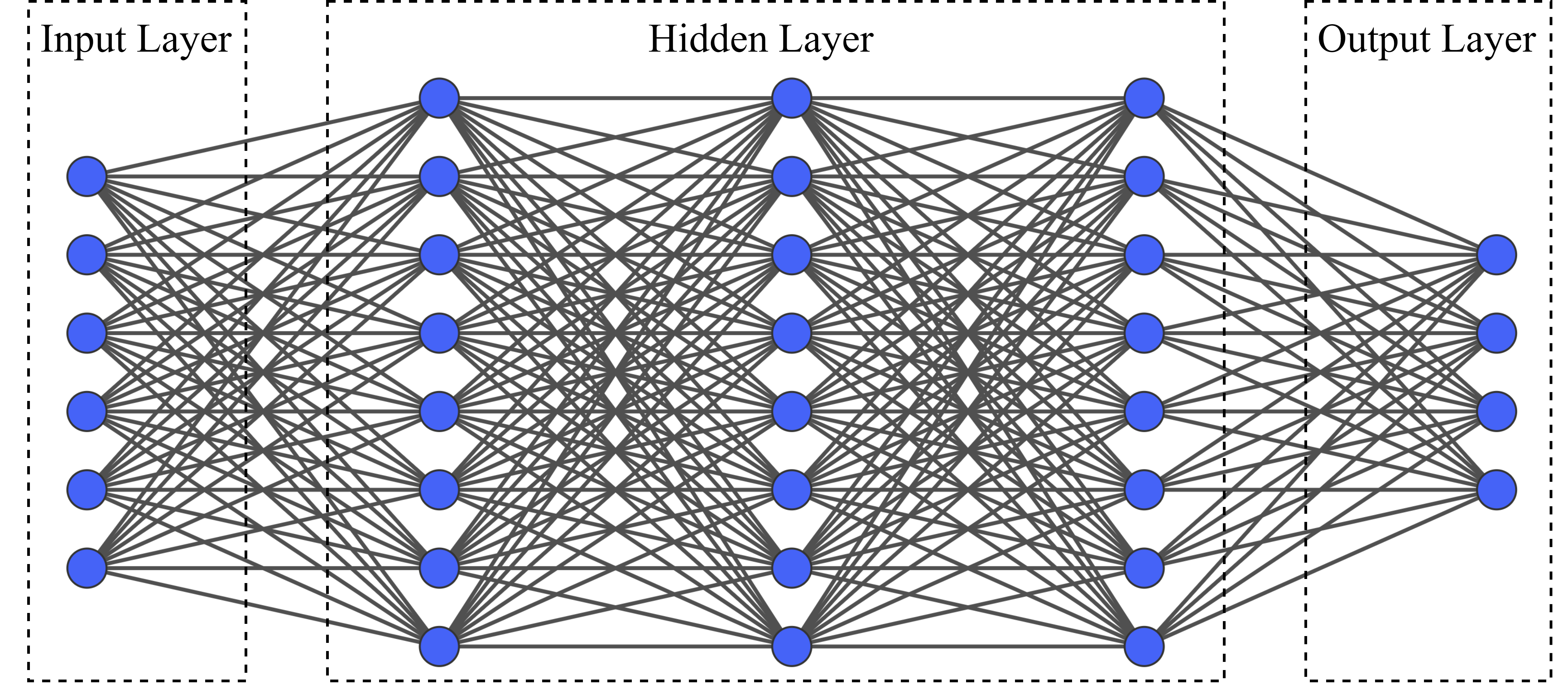}
\caption{Architecture of MLP, which has an input layer, an output layer, and three hidden layers.}
\label{fig_1}
\vspace {-0em}
\end{figure}

\emph{1) MLP}: MLP is a class of feed-forward neural network (FNN) with full connection between layers. The architecture of MLP is described in Fig. 1, which consists of an input layer, an output layer, and a lot of hidden layers. Each node in layers is a neuron, which sums up its weighted inputs along with a bias and propagates the result through a nonlinear activation function. The commonly nonlinear activation functions include the rectified linear unit (ReLU) function and sigmoid function, which are respectively defined as 
\begin{equation}
\begin{aligned}
{f_{\rm{sigmoid}}}(z) = {1 \mathord{\left/
 {\vphantom {1 {(1 + {e^{ - z}})}}} \right.
 \kern-\nulldelimiterspace} {(1 + {e^{ - z}})}},
 \end{aligned}
\end{equation}
\begin{equation}
\begin{aligned}
{f_{\rm{ReLU}}}(z) = \operatorname{max}\{ 0,z\}.
\end{aligned}
\end{equation}

\begin{figure}[!t]
\centering
\includegraphics[scale = 0.18]{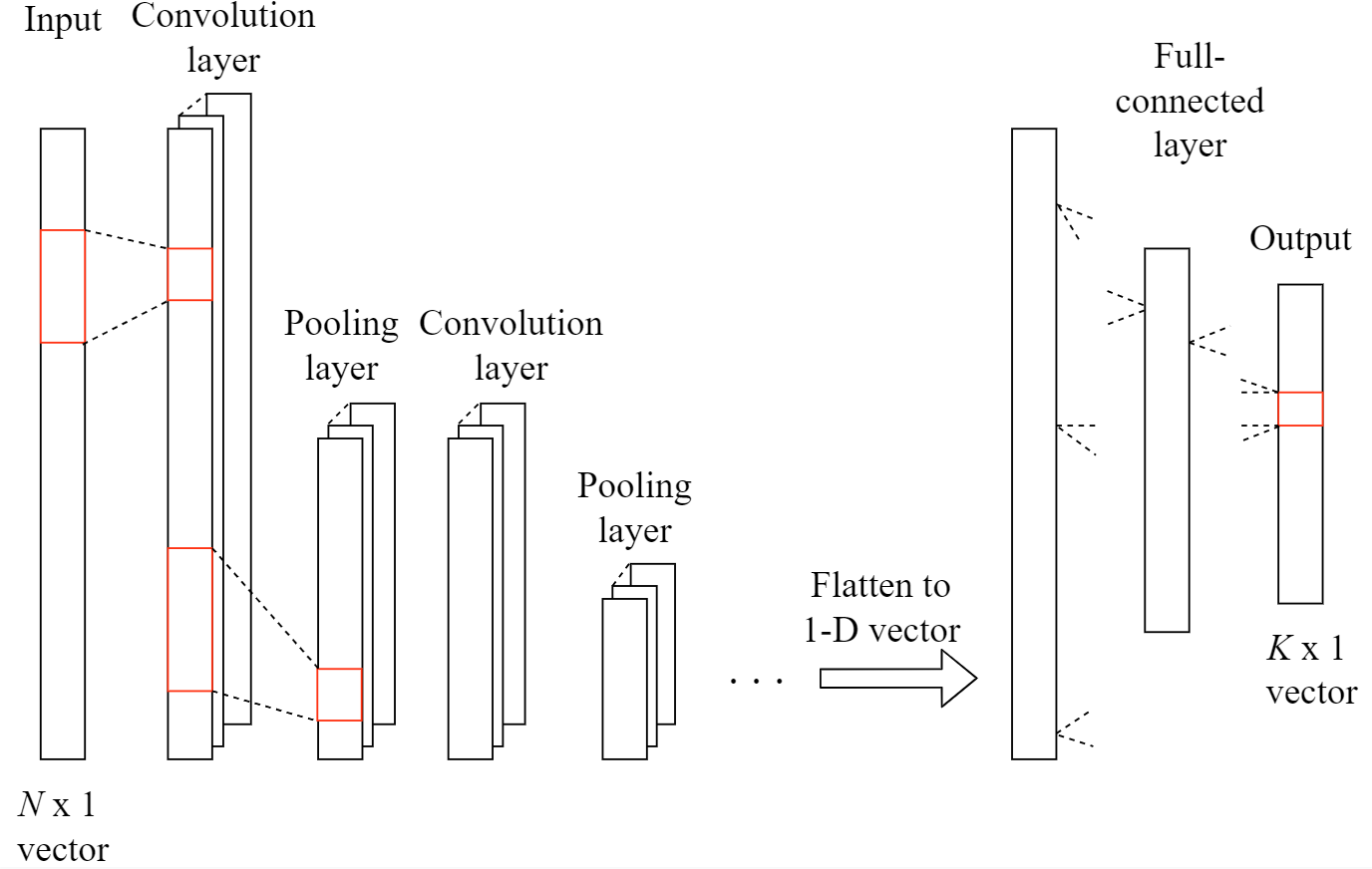}
\caption{Architecture of CNN, which consists of several convolution layers, pooling layers, and a fully-connected layer.}
\label{fig_2}
\vspace {0em}
\end{figure}

\emph{2) CNN}: 
CNN is a kind of FNN that has been successfully applied to computer vision tasks such as image classification, recognition, and segmentation. 
As depicted in Fig. 2, the hidden layers of classical CNN have two types: convolution layer and pooling layer. 
The convolution layer can be realized by a convolution kernel. 
The pooling layer is referred to as the sampling layer, whose typical applications include the MaxPooling layer and the MeanPooling layer \cite{23}. 
By introducing the convolution operation, the network parameters of CNN can be effectively reduced, thereby mitigating the risk of overfitting and simplifying the training process for DNN. 
The remarkable feature extraction capabilities of CNN have motivated researchers to design a CNN-based decoder. \par 

\begin{figure}[!t]
\centering
\includegraphics[scale = 0.21]{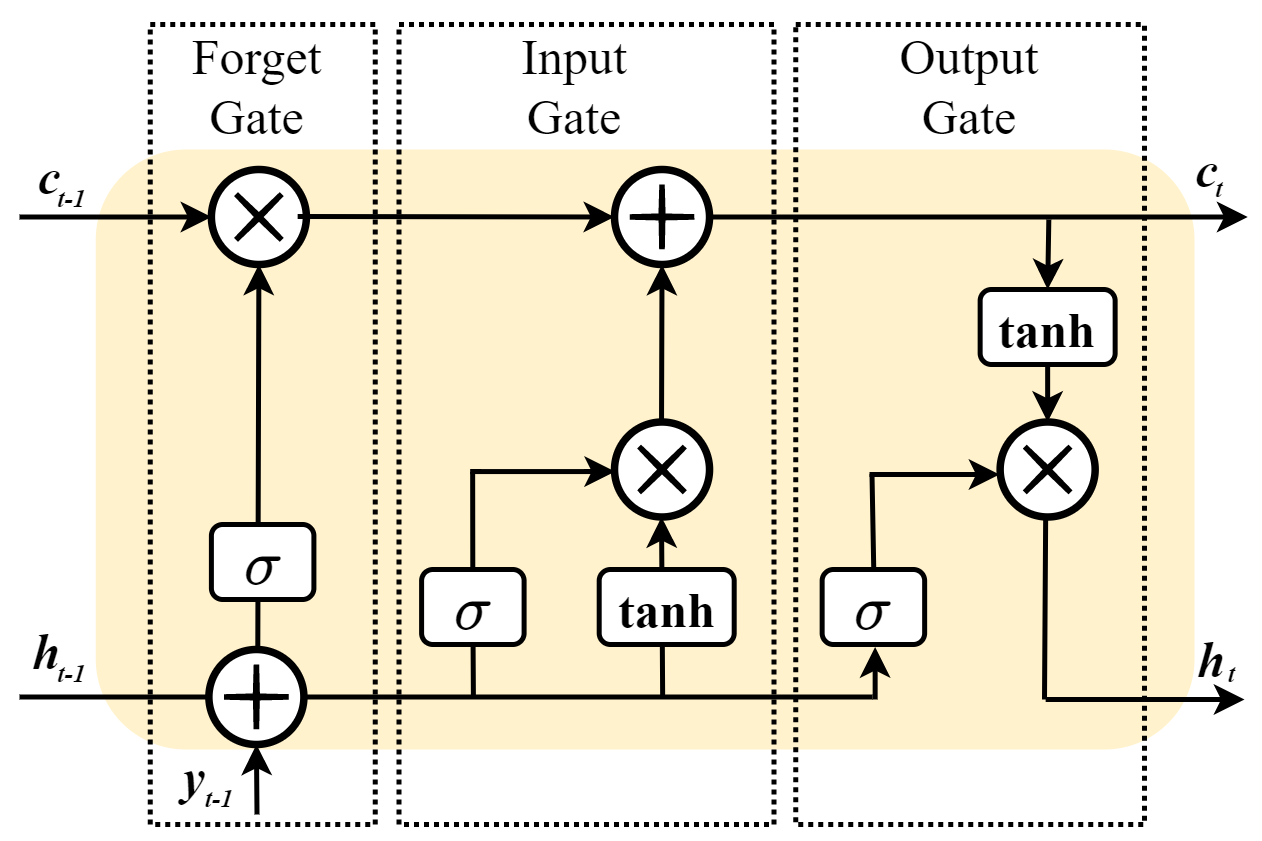}
\caption{Architecture of LSTM, where $\boldsymbol{h}_{t}$, $\boldsymbol{c}_{t}$, and $\boldsymbol{y}_{t}$ represent the hidden state, cell state, and input vector at time $t$, respectively.}
\label{fig_3}
\vspace {0em}
\end{figure}

\emph{3) RNN}: RNN is a type of neural network with a recurrent structure, which allows the previous states to influence the current output. Unlike FNNs, RNNs can use their internal memory to process arbitrary sequences of inputs. This feature makes them suitable for processing sequential data \cite{15}. Inspired by the remarkable performance of RNN on the time series task, many RNN-based decoders are proposed \cite{16}. Notably, general RNN suffers a serious vanishing gradient problem, which makes it difficult to be trained. Therefore, long short-term memory (LSTM) or gated recurrent unit (GRU) are commonly employed in practice. Consequently, we take LSTM as the representative of RNN. Fig. 3 shows the architecture of LSTM, which comprises three gates: forget gate, input gate, and output gate. The three gates are mainly used to control the information flow, thereby preventing issues associated with vanishing or exploding gradients. \par

\section{The Proposed DNN Decoders}

In this section, we first describe the system framework of DNN decoders. Then, the training process of DNN decoders is described in detail, and we provide a detailed explanation of some parameter settings, such as the $E_{b}/N_{0}$ of the training sample set. Finally, we present the network architecture and parameter design of MLP, CNN, and RNN, respectively. \par

\subsection{System Framework}

The system framework of the proposed DNN decoders is shown in Fig. 4. 
At the transmitter, $K$ bits information data $\boldsymbol{d}$ is first encoded into a $N$ bits codeword $\boldsymbol{x}$ with a PAC encoder. 
Then, $\boldsymbol{x}$ is mapped to modulated symbol vector $\boldsymbol{s}$ through binary phase shift keying (BPSK) modulation. The modulated symbol vector $\boldsymbol{s}$ is subsequently transmitted over a binary input additive white Gaussian noise (BI-AWGN) channel. The signal model for the received vector $\boldsymbol{y}$ can be written as 
\begin{equation}
\begin{aligned}
\boldsymbol{y}=\boldsymbol{s}+\boldsymbol{z},
 \end{aligned}
\end{equation}
where $\boldsymbol{z}$ represents the $1 \times N$ Gaussian noise vector. 
At the receiver, the estimated information bit vector $\boldsymbol{\hat{d}}$ is decoded from $\boldsymbol{y}$ by the DNN decoder. 
Therefore, the structure of DNN significantly affects the error-correction performance. 
In addition, the decoding loss function corrects the weights and biases in the neural network during the training phase. 
Since channel decoding can be regarded as a binary classification task, the  binary cross entropy (BCE) loss function is used as the decoding loss function, which can be formulated as
\begin{equation}
\begin{aligned}
{L_{\rm{BCE}}} =  - \frac{1}{K}\sum\limits_{i = 1}^K {\left( {\boldsymbol{d}[i]  \operatorname{log}\left( {\boldsymbol{p}[i]} \right) + \left( {1 - \boldsymbol{d}[i]} \right)  \operatorname{log}\left( {\boldsymbol{\bar{p}}[i]} \right)} \right)},
\end{aligned}
\end{equation}
where $\boldsymbol{p}[i]$ represents the probability value of $\boldsymbol{\hat{d}}[i]$, and $\boldsymbol{\bar{p}}[i]=1-\boldsymbol{p}[i]$. \par

\begin{figure}[!t]
\centering
\includegraphics[scale = 0.190]{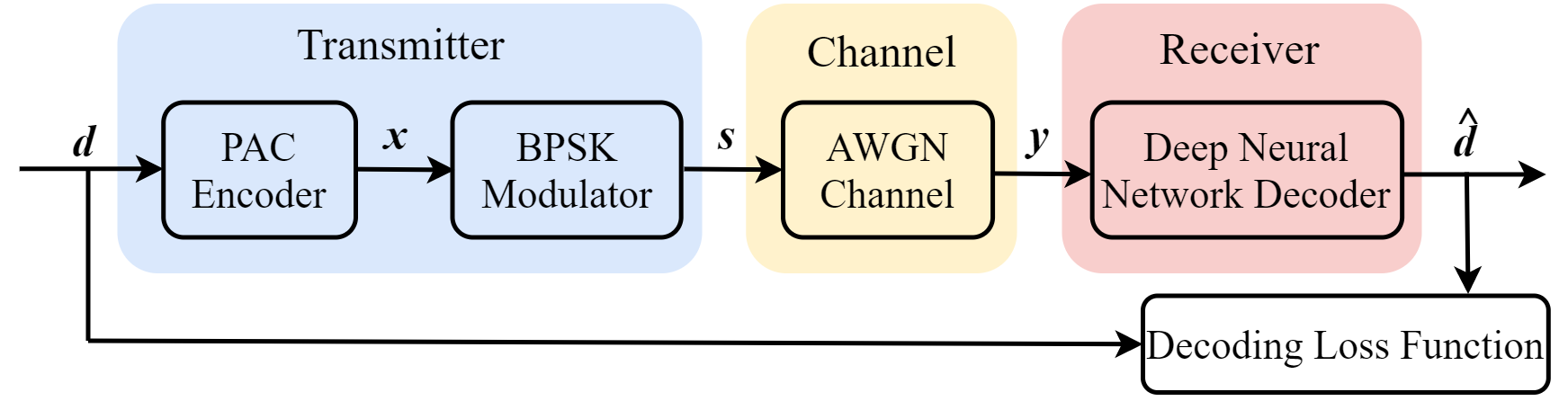}
\caption{The system framework of the proposed DNN decoders in this paper, where the decoding loss function corresponds to the equation (7).}
\label{fig_4}
\vspace {0em}
\end{figure}

\subsection{Training}

To find the optimal weights of the DNN, a training set with known input-output mappings is essential.
Compared with traditional DL scenarios such as image classification, the acquisition of training data for DNN decoders is simpler.
This is because the training samples are the $1 \times N$ received vector $\boldsymbol{y}$ and the labels of training samples are the $1 \times K$ information bit vector $\boldsymbol{d}$. 
However, to design the training phase of neural networks, which greatly influences the decoding performance of DNN decoders, we need to solve an important problem: finding the best $E_{b}/N_{0}$ to generate training samples. 
To solve this issue, we adopt the method in \cite{18} for the setting of training $E_{b}/N_{0}$ and define a performance metric known as the normalized validation error (NVE) as follows
\begin{equation}
\begin{aligned}
NVE({\rho _t}) = \frac{1}{{{S}}}\sum\limits_{o = 1}^{{S}} {\frac{{BER_{\rm{DNN}}({\rho _t},{\rho _{o}})}}{{BER_{\rm{Fano}}({\rho _{o}})}}},
\end{aligned}
\end{equation}
where ${\rho _{o}}$ denotes the $o$-th $E_{b}/N_{0}$ in the testing $E_{b}/N_{0}$ set, 
${\rho _{t}}$ represents the training $E_{b}/N_{0}$, 
$BER_{\rm{DNN}}({\rho _t},{\rho _{o}})$ is the bit error rate (BER) achieved by a DNN decoder (trains under ${\rho _{t}}$) at ${\rho _{o}}$, and $BER_{\rm{Fano}}({\rho _{o}})$ represents the BER performance of the Fano sequential decoder at ${\rho _{o}}$.
From equation (8), we know that the lower the NVE, the better the error-correction performance of the DNN decoder. 
In other words, there must be an optimal ${\rho_{t}}$ that enables the DNN decoder to achieve the minimum NVE. 
To get the optimal ${\rho_{t}}$, we train the three proposed DNN decoders with training sampling sets under different ${\rho_{t}}$ and evaluate the resulting NVE. 
As shown in Fig. 5, the trained DNN decoder is evaluated in terms of BER between the DNN decoder and the Fano sequential decoder via the Monte-Carlo (MC) simulation. 
It can be seen that there is the optimal training $E_{b}/N_{0}$ for different DNN decoders. 
Therefore, the training sampling sets under 0 dB, 4 dB, and 5 dB are chosen for MLP, CNN, and RNN decoders, respectively. \par

\begin{figure}[!t]
\centering
\includegraphics[scale = 0.45]{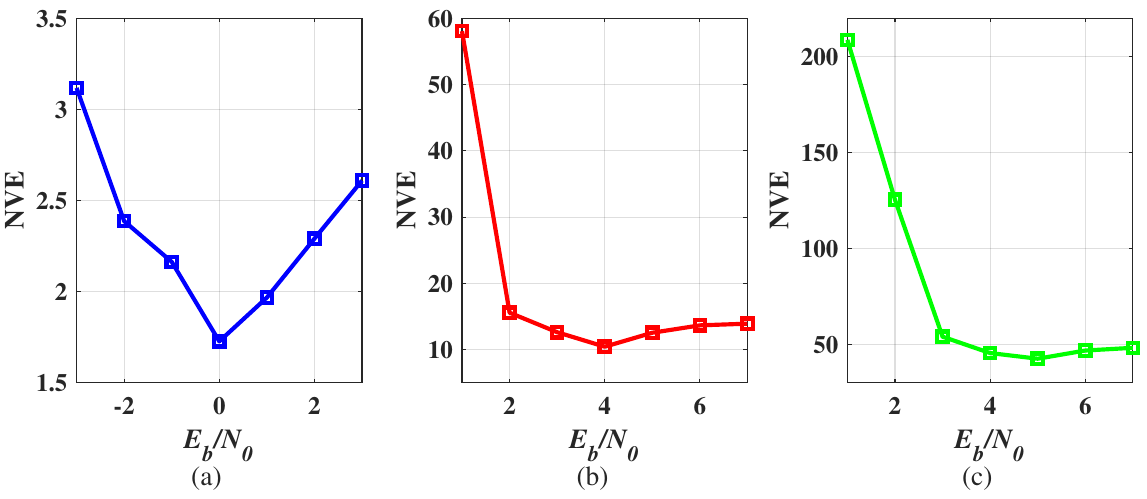}
\caption{NVE performance of DNN decoders with different training $E_{b}/N_{0}$ for PAC code (16, 8). (a) MLP decoder. (b) CNN decoder. (c) RNN decoder.}
\label{fig_3}
\vspace {0em}
\end{figure}

\begin{table}[!t]
\caption{The network parameter design of the three DNN decoders}
\begin{center}
\begin{tabular}{|c|c|c|c|}
\hline
Type             & MLP       & CNN                    & RNN          \\ \hline
Input Layer      & $N$       & $1 \times 1 \times N$  & $1 \times N$ \\ \hline
Hidden Layer 1   & Dense 128 & Conv1d $1 \times 128$  & LSTM 90     \\ \hline
Hidden Layer 2   & Dense 128 & Conv1d $128 \times 64$ & $\setminus$  \\ \hline
Hidden Layer 3   & Dense 128 & Conv1d $64 \times 16$  & $\setminus$  \\ \hline
Output Layer     & Dense $K$ & Dense $K$              & Dense $K$    \\ \hline
Total Parameters$^{1}$ & 35460     & 37652                  & 36364        \\ \hline
Total Parameters$^{2}$ & 37000     & 37848                  & 39608        \\ \hline
\end{tabular}%
\label{table_1}
\end{center}
\begin{tablenotes} 
\item $^{1}$ The number of parameters in DNN decoders for $N=8$ and $K=4$.
\item $^{2}$ The number of parameters in DNN decoders for $N=16$ and $K=8$. 
\end{tablenotes} 
\vspace {0em}
\end{table}

\subsection{Network Architecture and Parameter Design}
To ensure the number of parameters for different DNN decoders is within the same order of magnitude, we adjust the architecture of each DNN decoder.
The configurations of hidden layers are illustrated as follows: \par

\emph{1) MLP}: We use three fully connected layers to construct the MLP decoder, with each layer containing 128 nodes. 
Moreover, the ReLU function is selected for each hidden layer. \par

\emph{2) CNN}: 
We select three basic convolutional blocks as the three hidden layers of the CNN decoder. 
Each basic convolutional block consists of one convolutional layer, one MaxPooling layer, and the ReLU activation function. 
Notably, since the input of the decoder is the 1-D received signal vector, we adopt the convolutional layer with 1-D convolution instead of classical 2-D convolution.
The convolutional kernel size is set to 4 and the number of feature maps of the three convolutional layers are 128, 64, and 16, respectively.  
In addition, the MaxPooling layer with kernel size 2 and stride 2 is utilized between the adjacent convolutional layers. \par

\emph{3) RNN}: We adopt the LSTM to construct the RNN decoder. Due to the high complexity of LSTM, only one LSTM cell with a hidden dimension of 90 is used for the hidden layer. The input vector is sequentially fed symbol by symbol. After $N$ time steps, $K$ nodes are selected as the output nodes from the 90 nodes. \par

Since the dimension of the desired information vector is $K$, all three DNN decoders use a fully connected layer with $K$ nodes as the output layer. Moreover, the sigmoid function is selected as the activation function of the output layer. The detailed values of network parameters for the three DNN decoders are listed in Table I.  \par

\section{Experimental Results} 

\subsection{Experimental Environment}

\begin{table}[!t]
\caption{The training and testing hyper-parameter setting}
\begin{center}
\begin{tabular}{|c|c|}
\hline
Parameter              & Value             \\ \hline
Platform               & PyTorch             \\ \hline
Learning Rate          & 0.001             \\ \hline
Training Data          & $2^{20}$              \\ \hline
Batch Size             & 512               \\ \hline
Optimization Method    & Adam optimization \\ \hline
Testing $E_{b}/N_{0}$  & 0 dB - 6 dB       \\ \hline
\end{tabular}%
\label{table_1}
\end{center}
\vspace {0em}
\end{table}

We adopt the Reed-Muller polar (RM-polar) rate-profiling in \cite{8} to construct the PAC code, and the generator polynomial for the convolution operation in the PAC code is $\boldsymbol{g} = (1, 0, 1, 1, 0, 1, 1)$.
Experimental platforms comprise an NVIDIA 2080Ti GPU and an Intel Core i9-9900X CPU.
To select the relatively reasonable hyper-parameter of different neural networks, we conduct extensive experiments, and the final parameter setting is shown in Table II.
Notably, we have made parts of the source code of this paper available for reproducible research\footnote{https://github.com/daijingixn/Performance-Evaluation-of-PAC-Decoding-with-Deep-Neural-Networks}.
Moreover, error-correction experiments are conducted by utilizing BPSK modulation to send decoded codewords $\boldsymbol{x}$ on BI-AWGN channels. 
To ensure reliable results of error-correction performance experiments, a minimum of 500 errors are collected for each $E_{b}/N_{0}$ value. 

\subsection{Error-correction Performance Comparison}

\begin{figure*}[!t]
\centering
\includegraphics[scale = 0.71]{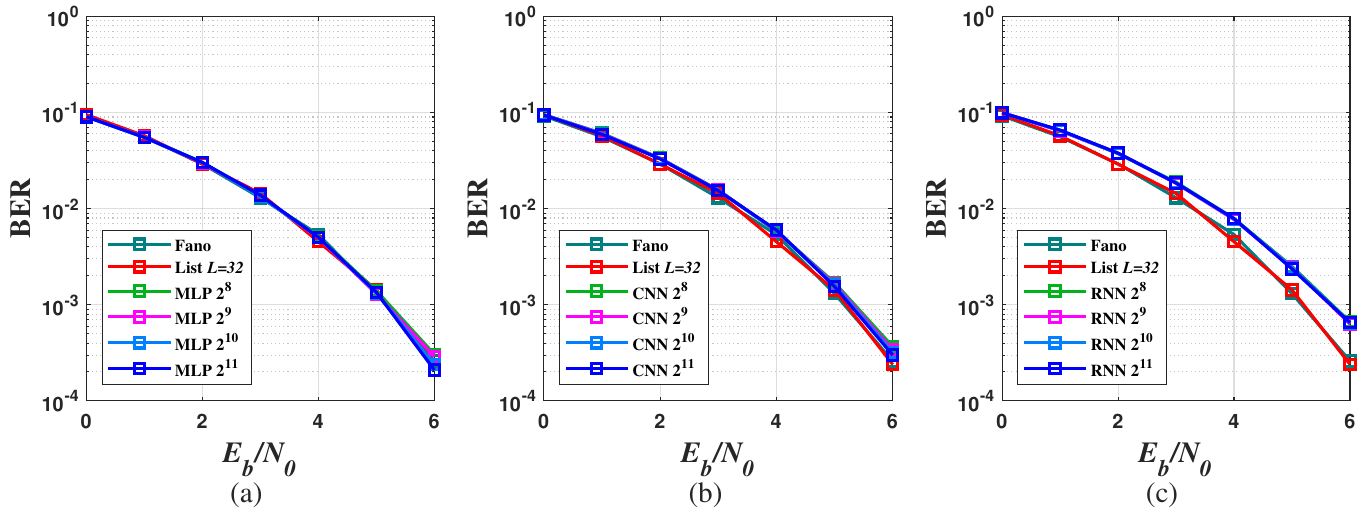}
\vspace {-0.5em}
\caption{BER performance of different DNN decoders ($N=8$ and $K=4$). (a) MLP decoder. (b) CNN decoder. (c) RNN decoder.}
\label{fig_7}
\vspace {0em}
\end{figure*}

\begin{figure*}[!t]
\centering
\includegraphics[scale = 0.71]{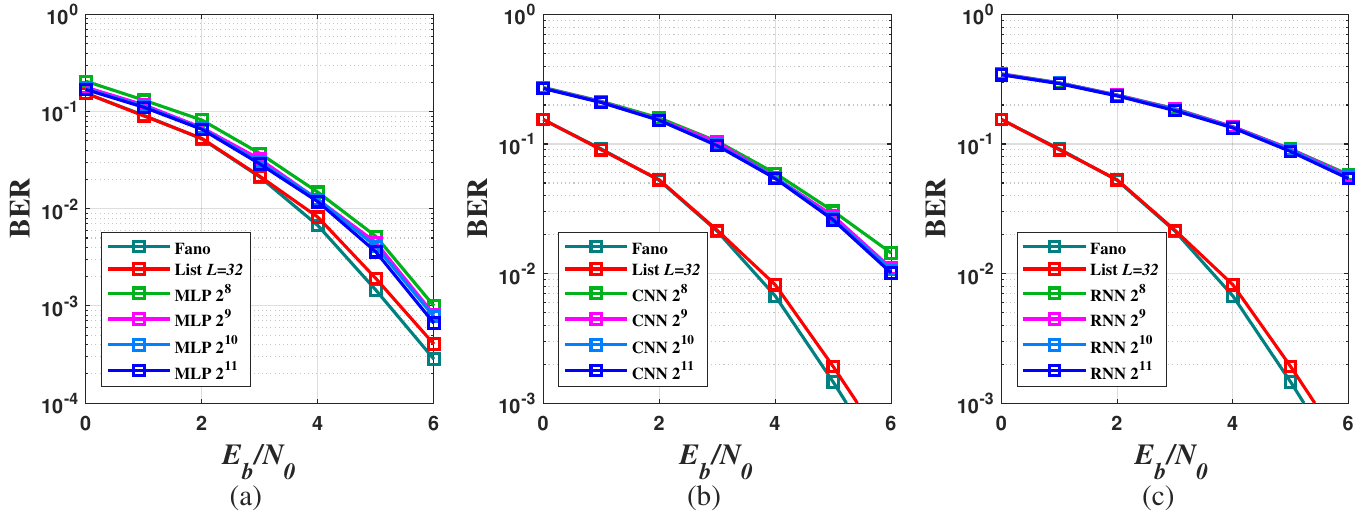}
\vspace {-0.5em}
\caption{BER performance of different DNN decoders ($N=16$ and $K=8$). (a) MLP decoder. (b) CNN decoder. (c) RNN decoder.}
\label{fig_7}
\vspace {0em}
\end{figure*}

In this section, we investigate the error-correction performance of the three proposed DNN decoders under various conditions. 
From Figs. 6 and 7, it is evident that the MLP decoder achieves the best performance among the three DNN decoders.
Compared to traditional decoders, DNN decoders offer the advantage of easy parallelization and are more suitable for hardware implementation. 
In the following part, experiments on the performance of the DNN decoder under different $N$, training epoch numbers, and input data are conducted.

\emph{1) Impact of code length $N$:}
We discuss the learning ability of different decoders by the saturation length $N_s$, which is the largest $N$ of DNN that can completely learn the encoding structure and noise characteristic.
It can be seen from Figs. 6 and 7 that the MLP and CNN decoders both show high learning ability on $N=8$, which is higher than the RNN decoder. 
For $N=16$, the MLP decoder shows the strongest learning ability.

\emph{2) Impact of epoch number:} Figs. 6 and 7 also show the influence of training epoch number on the BER performance of different DNN decoders.
For all DNN decoders, error-correction performance improves with a larger number of training epochs.
In other words, the performance gap between DNN decoders and traditional decoders decreases as the number of training epochs increases.
In addition, when the training epoch number is large, the error-correction performance of the DNN decoders gradually remains unchanged with the increasing epoch number.
\par

\emph{3) Impact of different input data:} 
In Fig. 8, we illustrate the influence of hard demodulation values, direct channel values, and channel LLR values as decoder input, respectively.
Notably, the types of input data for DNN decoders during the training and testing phases are the same.
MLP and CNN decoders show similar error-correction performance when using direct channel values and channel LLR values as inputs.
However, if hard demodulation values are used as input, only poor error-correction performance will be achieved.
In the case of the RNN decoder, using channel LLR values as input will achieve better results, while the other two input methods result in poorer performance.
\par

\subsection{Computational Time Comparison}

As shown in Fig. 9, we compare the computational time of different decoders. 
Noted, the computational time of the Fano sequential decoder is an average value from 0.0 dB to 6.0 dB.
Although the number of parameters in each DNN decoder is similar, each DNN decoder exhibits significant variations in computational time due to their distinct architectures.
Specifically, the MLP decoder requires the lowest time among all DNN decoders, which have 25.14\% and 94.02\% time reduction compared with Fano sequential and list decoders, respectively.
Moreover, we test the time of decoders for decoding one sample, which leads to the powerful parallel computing capability of GPU can not be fully leveraged.
Therefore, the computational time of GPUs is lower than CPUs.


\section{Conclusions} 
In this paper, we propose three types of DNN decoders, which build upon MLP, CNN, and RNN, respectively. 
Through experiments, we compare the performance of the three DNN decoders and find that MLP achieves the best decoding performance with similar parameter numbers and a shorter computational time.
Compared with traditional PAC decoding algorithms, the MLP decoder has lower latency while maintaining similar error-correction performance.
\par

\begin{figure}[!t]
\centering
\includegraphics[scale = 0.52]{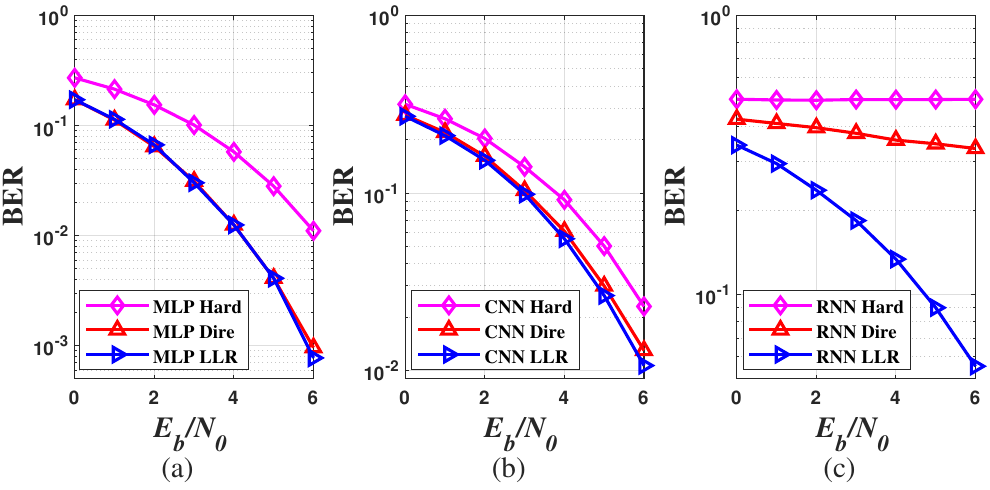}
\vspace {-0.5em}
\caption{BER performance of DNN decoders with different input data ($N=16$, $K=8$, and the training epoch is $2^{10}$). (a) MLP decoder. (b) CNN decoder. (c) RNN decoder.}
\label{fig_7}
\vspace {0em}
\end{figure}

\begin{figure}[!t]
\centering
\includegraphics[scale = 0.65]{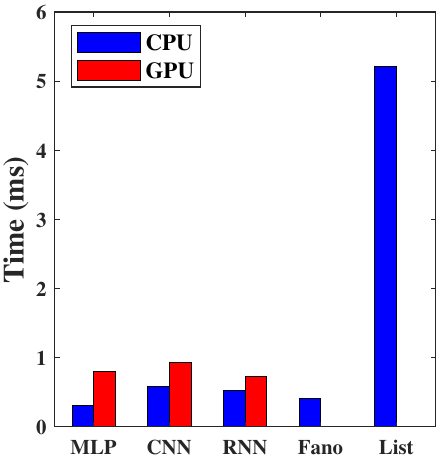}
\vspace {-0.5em}
\caption{Computational time of different decoders. 
}
\label{fig_9}
\vspace {0em}
\end{figure}

\section{Acknowledgment} 
This work is supported by the Fundamental Research Funds for the Central Universities under Grant CUC24BS07.

\balance 

\bibliography{Performance_Evaluation_of_PAC_Decoding_with_Deep_Neural_Networks}

\end{document}